# Observation of quadruple Weyl point in hybrid-Weyl phononic crystals

Li Luo[1*], Weiyin Deng[1*], Yating Yang[1], Mou Yan[1], Jiuyang Lu[1†], Xueqin Huang[1†], and Zhengyou Liu[2,3†]

[1]School of Physics and Optoelectronics and State Key Laboratory Luminescent Materials and Devices, South China University of Technology, Guangzhou 510640, China

[2]Key Laboratory of Artificial Micro- and Nanostructures of Ministry of Education and School of Physics and Technology, Wuhan University, Wuhan 430072, China

[3]Institute for Advanced Studies, Wuhan University, Wuhan 430072, China

*L.L. and W.D. contributed equally to this work

†Corresponding author. Email: phjylu@scut.edu.cn; phxqhuang@scut.edu.cn; zyliu@whu.edu.cn

The discovery of Weyl semimetals opens the door for searching topological semimetals in physical science. The Weyl points are generally recognized as conventional, quadratic, spin-1, and those of high topological charges. Here we report the observation of the quadruple Weyl point of charge 4, the highest topological charge a twofold degenerate node can carry. Besides the quadruple Weyl point, the phononic semimetal also hosts conventional, quadratic, and spin-1 Weyl points, which stands as a system with yet the richest types of Weyl points. The quadruple-helicoid surface states, specific to the quadruple Weyl point, are demonstrated. The finding of the high-charge Weyl point enriches the knowledge of Weyl semimetals and may stimulate related researches in other systems, such as photonic, mechanical and cold atom systems.

Searching new topological phases is of fundamental interest in condensed matter physics and has become a booming frontier in recent years. Besides topological insulators [1-4], the three-dimensional (3D) topological semimetals [5-11] have attracted great attentions since the prediction of Weyl semimetals [12-17]. In conventional Weyl semimetals, two bands linearly touch at the Fermi energy and constitute a Weyl point (WP) which possesses similar dispersions to the undiscovered Weyl particle in high energy physics [18-20]. WPs act as the monopoles of Berry curvatures in reciprocal space and carry nontrivial topological charges, which enables open surface Fermi arcs connecting the projections of opposite charged WPs. In addition to these conventional points that have their high energy counterparts, such as conventional WPs and Dirac points, different types of unconventional points have also been proposed, such as quadratic WPs [21-24].

Recently, twofold degenerate WPs of charge 4, dubbed as quadruple WPs, are theoretically predicted in spinless systems [25-27]. The existence of the quadruple WPs needs cubic symmetries, which are point groups $O$ and $T$ with time-reversal symmetry [26]. With the cubic symmetry, the dispersions of the quadruple WPs have specific cubic profile in the (111) direction and quadratic in all other directions. The quadruple WPs have the largest topological charge a twofold degenerate point can host [27], in which, according to bulk-boundary correspondence, quadruple-helicoid surface arc states can be anticipated.

In general, the precise control of the crystalline symmetry can facilitate the observation of these symmetry-protected WPs. Phononic crystal (PC), for its macroscopic and controllable features, provides a well-performed and feasible platform, where many acoustic analogues of topological insulators, including Chern topological insulators [28-30], non-Hermitian topological insulators [31, 32] and higher-order topological insulators [33-35] have been realized. More importantly, PC is a naturally spinless systems, which makes it possible to observe the quadruple WPs. Recently, chiral structures have been introduced into PCs to observe the acoustic Weyl and Dirac semimetals [15, 36-39]. However, the existing PCs are limited to structures with chiral couplings only in one direction, which have a lower crystalline symmetry and can merely host WPs of charges less than four.

In this work, we introduce chiral couplings in all three directions and construct a PC with the space group symmetry $P432$. We find that this PC hosts multiple WPs in the Brillouin zone (BZ), including both twofold and threefold WPs carrying different topological charges. All these WPs are tabulated in Table I. Among them, a twofold quadruple WP with a topological charge 4 is observed at the BZ center, which has not been experimentally realized so far. We elaborately design the PC and achieve the quadruple WP and other WPs. The measured surface arcs confirm the topological properties of these WPs.

Our 3D printing sample is arranged in a cubic lattice with its unit cell shown in Fig. 1(a). Considering the periodicity, only three inequivalent cavities, connected by straight tubes, exist in one unit cell. This structure can be well mapped to a tight-binding lattice model shown schematically in Fig. 1(b), where cavities correspond to grid nodes and tubes map to hoppings (with strength $t_0$) between different nodes. Setting the lattice constant to be one, the Hamiltonian takes the form

$$H = \Sigma_i d_i \lambda_i = \begin{pmatrix} 0 & d_1 - id_2 & d_4 - id_5 \\ d_1 + id_2 & 0 & d_6 - id_7 \\ d_4 + id_5 & d_6 + id_7 & 0 \end{pmatrix}, \quad (1)$$

where $\lambda_i$ are the $3 \times 3$ Gell-Mann matrices with the coefficients $d_1 = 4t_0 \cos\frac{k_x}{2} \cos\frac{k_y}{2} \cos k_z$, $d_2 = 4t_0 \sin\frac{k_x}{2} \sin\frac{k_y}{2} \sin k_z$, $d_4 = 4t_0 \cos\frac{k_x}{2} \cos k_y \cos\frac{k_z}{2}$, $d_5 = -4t_0 \sin\frac{k_x}{2} \sin k_y \sin\frac{k_z}{2}$, $d_6 = 4t_0 \cos k_x \cos\frac{k_y}{2} \cos\frac{k_z}{2}$, and $d_7 = 4t_0 \sin k_x \sin\frac{k_y}{2} \sin\frac{k_z}{2}$. From Eq. (1), the band structures along high symmetry lines [the red lines in Fig. 1(d)] are presented by the red dashed curves in Fig. 1(c). Note that the hopping strength $t_0$ (typically chosen as $-1$) is the solely adjustable parameter, which only scales the band structure and does not change the band crossings. On the other hand, due to the constraints of the space group symmetry ($P432$), the simulated dispersions of the PC [the black curves in Fig. 1(c)] are extremely stable and coincide with the lattice model, which confirms the effectiveness of the mapping between the PC and the lattice model.

This chiral structure possesses a series of stable band crossings, classified as different kinds of WPs and labelled by different colors in Fig. 1(c). Specifically, the band crossings marked by orange, yellow, and gray balls are conventional WPs with linear dispersions in all the directions and topological charges $\pm 1$. They share the standard low-energy Hamiltonian $\delta H = \Sigma_i k_i \sigma_i$, $i = x, y, z$, where $k_i$ are the wavevectors and $\sigma_i$ are the Pauli matrices. Note that to characterize a general two-dimensional Hamiltonian $H = \Sigma_i d_i(\mathbf{k}) \sigma_i$, one can define a normalized vector $\hat{d} = \vec{d}/|\vec{d}|$, which coincides with the expectation value of $\sigma_i$ with respect to the states in the upper band. In this way, the normalized vector can be viewed as a pseudospin of the WP. For the conventional WP, its pseudospin texture over a sphere in the reciprocal space is shown in the left panel of Fig. 1(e). With the hedgehog-like pseudospin distribution, an integral wrapping number is represented by the number of times the pseudospin sweeps the sphere, which equals to the topological charge of the WP and provides a geometric perspective on the charge. The pseudospin definition can be extended to the WPs of spin-1 labeled by the purple and magenta balls in Fig. 1(c) (see Supplementary Materials [40]).

The WP marked by the blue ball in Fig. 1(c) is quadratic, which carries a topological charge of 2. Its dispersion is quadratic in two of the three mutually perpendicular directions and linear in the other one (see Supplementary Materials [40]). For the lattice model, the Hamiltonian coefficients are $d_x = \sqrt{3}\,(k_x^2 - k_y^2)/4$, $d_y = 3\,k_x k_y/4$, and $d_z = 2\sqrt{3}k_z + (k_x^2 + k_y^2)/4$, and the pseudospin distribution is accordingly plotted in the middle panel of Fig. 1(e), where from the top view of the sphere, the pseudospin rotates $4\pi$ around the equator [the red circle in Fig. 1(e)].

Most importantly, our sample hosts a twofold degenerate point at the center of the BZ [the green ball in Fig. 1(c)], which is a quadruple WP with $C = 4$, the highest topological charge of a twofold degenerate node in spinless systems [25-27]. Reduced from Eq. (1), the low-energy Hamiltonian near this quadruple WP is

$$\delta H = -k^2 \sigma_0 + \frac{\sqrt{3}}{2}(k_x^2 - k_y^2)\sigma_x + \left(\frac{1}{2}k_x^2 + \frac{1}{2}k_y^2 - k_z^2\right)\sigma_y + \sqrt{3}k_x k_y k_z \sigma_z, \quad (2)$$

where $k^2 = k_x^2 + k_y^2 + k_z^2$ and $\sigma_0$ is a two-by-two identity matrix. For this quadruple WP, its pseudospin texture [the right panel of Fig. 1(e)] is quite different from the above two cases and overlays the sphere four times. The robustness of the quadruple WP lies on the protection of the chiral cubic symmetry, which possesses a two-dimensional irreducible representation at the $\Gamma$ point and embody the specific form of the Hamiltonian. Supplementary Materials [40] provides the derivation of Eq. (2), where the topological charge is analytically proved to be four. The calculated dispersions near the quadruple WPs are shown in Fig. 1(f). This log-log plot clearly exhibits the cubic dispersion along (111) direction and quadratic dispersions along the other two directions, which is the prominent feature of the quadruple WP.

These different kinds of WPs can be demonstrated by measuring the projected bulk dispersions. The sample photo is shown in Fig. 2(a). To measure the projected bulk dispersion, a headphone with a diameter 6 mm is embedded in a cavity at the center beneath the YZ surface as the excitation source; then a microphone probe scans on the YZ plane at a distance of 8 mm above the source and detects sound signals emitting from the holes; the surface sound field distribution is finally recorded by network analyzer (Keysight E5061B). The experimental setup is illustrated in Fig. 2(b). The corresponding projected bulk dispersions can be obtained by performing spatial Fourier transformation on the scanned acoustic signals. Figure 2(c) exbibits the projected bulk dispersions along the high symmetry lines $\bar{\Gamma}$-$\bar{M}$-$\bar{R}$-$\bar{\Gamma}$ in the $k_y$-$k_z$ plane [the blue surface in Fig. 1(d)]. The simulated projected bulk dispersions are plotted as grey curves, in which different kinds of WPs are represented by the crossings of the dispersions.

A specific feature of Weyl semimetals is the open Fermi arcs, that is, the dispersions of surface states start from and end up to the projections of WPs on the surface. Since our structure involves rich and diverse WPs, especially the quadruple one, it is worth to investigate the surface state dispersions in the sample. To measure the dispersions, acoustic fields in each hole on the boundary are detected. The setup is similar to that illustrated in Fig. 2(b), but the boundary condition here are acoustically rigid to confine the acoustic surface states. Therefore, resin covers are used to seal all the holes on the YZ surface, and the probe is embedded in a movable cover to scan the acoustic fields cell by cell. The measured surface state dispersions can be achieved by Fourier transforming the scanned acoustic signals on the surface.

Here we exemplify the surface state dispersions in two cases, one relates to the quadratic WP and the other to the quadruple one. For the first case, the simulated iso-frequency contours are plotted at 2.60 kHz, between the frequencies of the blue and yellow WPs. As shown in Fig. 3(a), eight arcs exist in the projected BZ, connecting the projections of these WPs. Specifically, since two overlapped WPs (of a topological charge $-1$) are projected to the same yellow point with a total charge $(-1) \times 2$, two arcs emanate from each yellow point and tends to the projections of the WPs (blue) at the BZ center and near the BZ boundary, respectively. Similarly, because two quadratic WPs (blue) are projected to the BZ center and each carries a topological charge $+2$, four arcs converge to the BZ center. For the projection of the quadratic WP near the BZ boundary, there is only one Fermi arc links to it in this contour, since the other Fermi arc appears at a higher frequency connecting to the projections of a spin-1 WP at 2.99 kHz (see Supplementary Materials [40]). For the second case, the iso-frequency contours at the frequency of 3.55 kHz are shown in Fig. 3(b), where four open arcs connecting the projections of the green and grey WPs are measured. These arcs originate from the quadruple WP (green) with $C = 4$, and terminate into four conventional ones (gray) with $C = -4$. The measured results agree well with the corresponding numerical iso-frequency curves [marked in black in Fig. 3(b)].

In fact, both the number and sign of the topological charges of the WPs can be determined simultaneously by tracking the number and trend of the chiral gapless surface states arising from the projected bulk bands [5, 24, 41]. A practically and widely used approach is to choose a cylinder in the 3D BZ with its baseline on the projected BZ, as illustrated in Figs. 3(c) and 3(d). Because of the bulk-edge correspondence, the surface states are generated by the charges of the WP enclosed by the cylinder. Specifically, as shown in Fig. 3(c), the cylinder encloses two yellow balls, and the dispersions in Fig. 3(e) shows two surface states going through the bulk band gap from high to low frequencies, indicating that each WP marked in yellow is of a topological charge $C = -1$. Following the similar procedure, we can determine the charge of the quadruple WP. Using the cylinder to enclose the green ball [Fig. 3(d)], there exist 4 gapless surface states, known as the quadruple-helicoid surface states, traversing the projected bulk band gap chirally from low to high frequencies, as shown consistently by both measured and simulated dispersions in Fig. 3(f). This provides direct evidence for the existence of quadruple WP of a topological charge $C = 4$, which is the highest topological charge observed ever in experiment.

In conclusion, we have realized a multi-Weyl PC. Benefited from the introduced chiral couplings in all three directions, the PC is robust to host multiple WPs and can be well captured by a simple effective model. In particular, the quadruple WP with topological charge $C = 4$ and corresponding quadruple-helicoid surface states have been clearly observed. This firmly advances the research of semimetal physics, especially in photonic and phononic crystals. The quadruple WPs may serve new opportunity for exploring exotic topological phases, such as higher-order phases [42-44]. This PC can further possess peculiar properties and may produce applications in surface state transports enhanced by multiple WPs, which can stimulate related research in other systems.

**Acknowledgement**

The authors thank Feng Li for the helpful discussions about the experimental sample. This work was supported by the National Natural Science Foundation of China (11890701, 11974120, 11974005, 12074128), the National Key R&D Program of China (2018YFA0305800), the Guangdong Basic and Applied Basic Research Foundation (2019B151502012, 2021B1515020086, and 2021A1515010347), and the Guangdong Innovative and Entrepreneurial Research Team Program (No. 2016ZT06C594).

**References**

[1] M. Z. Hasan, and C. L. Kane, Colloquium: Topological insulators, Rev. Mod. Phys. **82**, 3045 (2010).

[2] X. L. Qi, and S. C. Zhang, Topological insulators and superconductors, Rev. Mod. Phys. **83**, 1057 (2011).

[3] L. Fu, Topological crystalline insulators, Phys. Rev. Lett. **106**, 106802 (2011).

[4] A. Bansil, H. Lin, and T. Das, Colloquium: Topological band theory, Rev. Mod. Phys. **88**, 021004 (2016).

[5] X. Wan, A. M. Turner, A. Vishwanath, and S. Y. Savrasov, Topological semimetal and Fermi-arc surface states in the electronic structure of pyrochlore iridates, Phys. Rev. B **83**, 205101 (2011).

[6] C. Fang, M. J. Gilbert, X. Dai, and B. A. Bernevig, Multi-Weyl topological semimetals stabilized by point group symmetry, Phys. Rev. Lett. **108**, 266802 (2012).

[7] B. Q. Lv, H. M. Weng, B. B. Fu, X. P. Wang, H. Miao, J. Ma, P. Richard, X. C. Huang, L. X. Zhao, G. F. Chen, Z. Fang, X. Dai, T. Qian, and H. Ding, Experimental discovery of Weyl semimetal TaAs, Phys. Rev. X **5**, 031013 (2015).

[8] A. A. Burkov, Topological semimetals, Nat. Mater. **15**, 1145-1148 (2016).

[9] C. Fang, L. Lu, J. Liu, and L. Fu, Topological semimetals with helicoid surface states, Nat. Phys. **12**, 936-941 (2016).

[10] N. P. Armitage, E. J. Mele and A. Vishwanath, Weyl and Dirac semimetals in three-dimensional solids, Rev. Mod. Phys. **90**, 015001 (2018).

[11] N. B. M. Schröter, D. Pei, M. G. Vergniory, Y. Sun, K. Manna, F. De. Juan, J. A. Krieger, V. Süss, M. Schmidt, P. Dudin, B. Bradlyn, T. K. Kim, T. Schmitt, C.

Cacho, C. Felser, V. N. Strocov, and Y. Chen, Chiral topological semimetal with multifold band crossings and long Fermi arcs, Nat. Phys. **15**, 759-765 (2019).

[12] S.-Y. Xu, I. Belopolsk, N. Alidoust, M. Neupane, G. Bian, C. Zhang, R. Sankar, G. Chang, Z. Yuan, C.-C. Lee, S.-M. Huang, H. Zheng, J. Ma, D. S. Sanchez, B. Wang, A. Bansil, F. Chou, P. P. Shibayev, H. Lin, S. Jia, and M. Z. Hasan, Discovery of a Weyl fermion semimetal and topological Fermi arcs, Science **349**, 613-617 (2015).

[13] S.-Y. Xu, N. Alidoust, I. Belopolski, Z. Yuan, G. Bian, T.-R. Chang, H. Zheng, V. N. Strocov, D. S. Sanchez, G. Chang, C. Zhang, D. Mou, Y. Wu, L. Huang, C.-C. Lee, S.-M. Huang, B. Wang, A. Bansil, H.-T. Jeng, T. Neupert, A. Kaminski, H. Lin, S. Jia, and M. Z. Hasan, Discovery of a Weyl fermion state with Fermi arcs in niobium arsenide, Nat. Phys. **11**, 748-754 (2015).

[14] K. Deng, G. Wan, P. Deng, K. Zhang, S. Ding, E. Wang, M. Yan, H. Huang, H. Zhang, Z. Xu, J. Denlinger, A. Fedorov, H. Yang, W. Duan, H. Yao, Y. Wu, S. Fan, H. Zhang, X. Chen, and S. Zhou, Experimental observation of topological Fermi arcs in type-II Weyl semimetal $MoTe_2$, Nat. Phys. **12**, 1105-1110 (2016).

[15] F. Li, X. Huang, J. Lu, J. Ma, and Z. Liu, Weyl points and Fermi arcs in chiral phononic crystal, Nat. Phys. **14**, 30-34 (2018).

[16] B. Xie, H. Liu, H. Cheng, Z. Liu, S. Chen, and J. Tian, Experimental realization of type-II Weyl points and Fermi arcs in Phononic crystal, Phys. Rev. Lett. **122**, 104302 (2019).

[17] X. Huang, W. Deng, F. Li, J. Lu, and Z. Liu, Ideal type-II Weyl phase and topological transition in phononic crystals, Phys. Rev Lett. **124**, 206802 (2020).

[18] L. X. Yang, Z. K. Liu, Y. Sun, H. Peng, H. F. Yang, T. Zhang, B. Zhou, Y. Zhang, Y. F. Guo, M. Rahn, D. Prabhakaran, Z. Hussain, S.-K. Mo, C. Felser, B. Yan, and Y. L. Chen, Weyl semimetal phase in the non-centrosymmetric compound TaAs, Nat. Phys. **11**, 728-732 (2015).

[19] W.-J. Chen, M. Xiao, and C.T. Chan, Photonic crystals possessing multiple Weyl points and the experimental observation of robust surface states, Nat. commun. **7**, 1-10 (2016).

[20] D. F. Liu, A. J. Liang, K. E. Liu, Q. N. Xu, Y. W. Li, C. Chen, D. Pei, W. J. Shi, S. K. Mo, P. Dudin, T. Kim, C. Cacho, G. Li, Y. Sun, L. X. Yang, Z. K. Liu, S. S. P. Parkin, C. Felser, and Y. L. Chen, Magnetic Weyl semimetal phase in a Kagomé crystal, Phys. Rev. Lett. **365**, 1282-1285 (2019).

[21] M.-L. Chang, M. Xiao, W.-J. Chen, and C. T. Chan, Multiple Weyl points and the sign change of their topological charges in woodpile photonic crystals, Phys. Rev. B **95** 125136 (2017).

[22] H. He, C. Qiu, L. Ye, X. Cai, X. Fan, M. Ke, F. Zhang, and Z. Liu, Topological negative refraction of surface acoustic waves in a Weyl phononic crystal, Nature **50**, 61-64 (2018).

[23] T. Zhang, Z. Song, A. Alexandradinata, H. Weng, C. Fang, L. Lu, and Z. Fang, Double-Weyl phonons in transition-metal monosilicides, Phys. Rev. Lett. **120**, 016401 (2018).

[24] H. He, C. Qiu, X. Cai, M. Xiao, M. Ke, F. Zhang, and Z. Liu, Observation of quadratic Weyl points and double-helicoid arcs, Nat. Commun. **11**, 1-6 (2020).

[25] T. Zhang, R. Takahashi, C. Fang, and S. Murakami, Twofold quadruple Weyl nodes in chiral cubic crystals, Phys. Rev. B **102**, 125148 (2020).

[26] C. Cui, X.-P. Li, D.-S. Ma, Z.-M Yu, and Y. Yao, Charge-four Weyl point: minimum lattice model and chirality-dependent properties. Phys. Rev. B **104**, 075115 (2021).

[27] Z.-M Yu, Z. Zhang, G.-B. Liu, W. Wu, X.-P. Li, R.-W. Zhang, S. A. Yang, and Y. Yao, Encyclopedia of emergent particles in three-dimensional crystals. arXiv preprint arXiv:2102.01517 (2021).

[28] J. Lu, C. Qiu, W. Deng, X. Huang, F. Li, F. Zhang, S. Chen, and Z. Liu, Valley topological phases in bilayer sonic crystals. Phys. Rev. Lett. **120**, 116802 (2018).

[29] W. Deng, X. Huang, J. Lu, V. Peri, F. Li, S. D. Huber, and Z. Liu, Acoustic spin-Chern insulator induced by synthetic spin-orbit coupling with spin conservation breaking, Nat. Commun. **11**, 1-7 (2020).

[30] Z. Zhu, M. Yan, J. Pan, Y. Yang, W. Deng, J. Lu, X. Huang, and Z. Liu, Acoustic Valley Spin Chern Insulators, Phys. Rev. Appl. **16**, 014058 (2021).

[31] C. Yuce, Edge states at the interface of non-Hermitian systems, Phys. Rev. A **97**, 042118 (2018).

[32] B. Höckendorf, A. Alvermann, and H. Fehske, Non-Hermitian boundary state engineering in anomalous Floquet topological insulators, Phys. Rev. Lett. **123**, 190403 (2019).

[33] H. Xue, Y. Yang, F. Gao, Y. Chong, and B. Zhang, Acoustic higher-order topological insulators on kagome lattice, Nat. Mater. **18**, 108-112 (2018).

[34] H. Xue, Y. Ge, H.-X. Sun, Q. Wang, D. Jia, Y.-J. Guan, S.-Q. Yuan, Y. Chong, and B. Zhang, Observation of an acoustic octupole topological insulator, Nat. Commun. **11**, 1-6 (2020).

[35] Z.-G. Chen, W. Zhu, Y. Tan, L. Wang, and G. Ma, Acoustic realization of a four-dimensional higher-order Chern insulator and boundary-modes engineering, Phys. Rev. X **11**, 011016 (2021).

[36] Y. Qi, C. Qiu, M. Xiao, H. He, M. Ke, and Z. Liu, Acoustic realization of quadrupole topological insulators, Phys. Rev. Lett. **124**, 206601 (2020).

[37] M. Xiao, W.-J. Chen, W.-Y. He, and C. T. Chan, Synthetic gauge flux and Weyl Points in acoustic systems, Nat. Phys. **11**, 920-924 (2015).

[38] H. Cheng, Y. Sha, R. Liu, C. Fang, and L. Lu, Discovering topological surface states of Dirac points, Phys. Rev Lett. **124**, 104301 (2020).

[39] W. Deng, X. Huang, J. Lu, F. Li, J. Ma, S. Chen, and Z. Liu, Acoustic spin-1 Weyl semimetal, SCIENCE CHINA Physics, Mechanics & Astronomy. **63**, 1-6 (2020).

[40] See Supplemental Material for further details.

[41] S. Vaidya, J. Noh, A. Cerjan, C. Jörg, G. V. Freymann, and M. C. Rechtsman, Observation of a Charge-2 Photonic Weyl Point in the Infrared, Phys. Rev Lett. **125**, 253902 (2020).

[42] Q. Wei, X. Zhang, W. Deng, J. Lu, X. Huang, M. Yan, G. Chen, Z. Liu, and S. Jia, Higher-order topological semimetal in acoustic crystals, Nat. Mater. **20**, 812 (2021).

[43] L. Luo, H.-X. Wang, Z.-K. Lin, B. Jiang, Y. Wu, F. Li, and J.-H. Jiang, Observation of a phononic higher-order Weyl semimetal, Nat. Mater. **20**, 794 (2021).

[44] H. Qiu, M. Xiao, F. Zhang, and C. Qiu, Higher-order Dirac sonic crystal, Phys. Rev Lett. **127**, 146601 (2021).

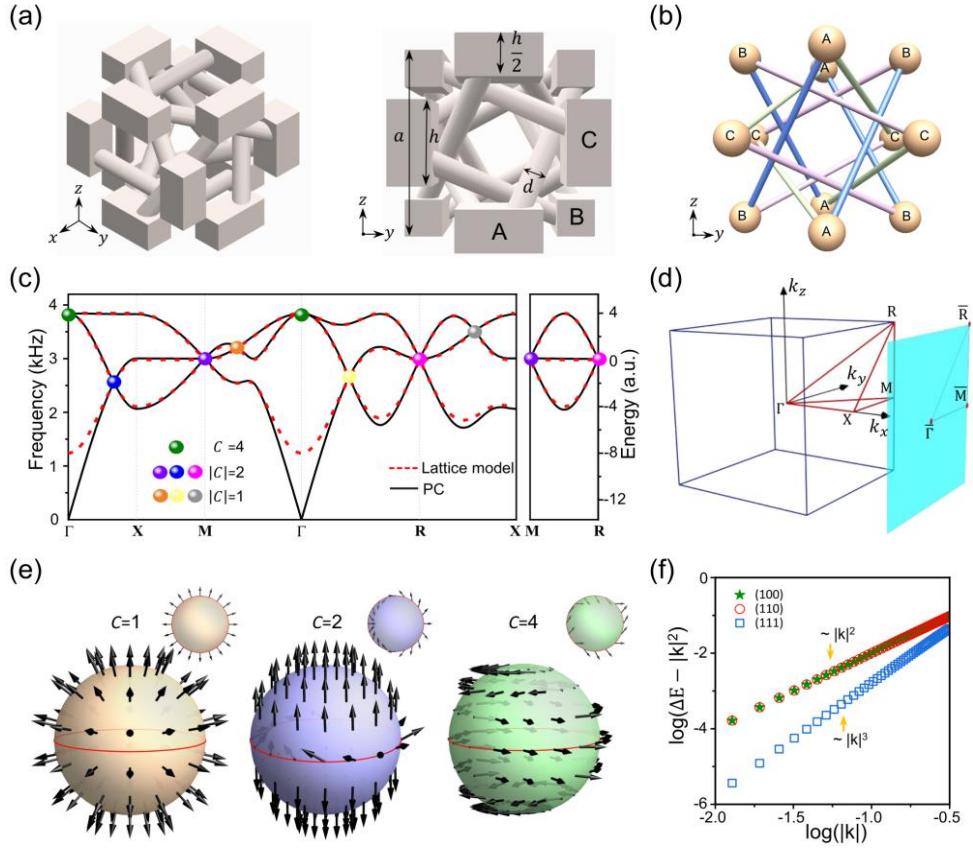

FIG. 1. (a) Side and front views of a unit cell of the chiral PC, where the lattice constant $a = 20$ mm, the side length of cubic cavity $h = 7.92$ mm, and the tube diameter $d = 2.26$ mm. (b) Schematic of the lattice model corresponding to the PC. Hoppings along different directions are marked in three different colors for clarity. (c) Bulk dispersions of the PC (black solid curves) and the lattice model (red dashed curves). (d) The first BZ and the projected BZ on the $k_y$-$k_z$ plane, where high symmetry points and lines are marked out. (e) Pseudospin textures for the WPs of charges 1, 2 and 4, respectively. Insets: Top views of the pseudospins at the equators (red circles). (f) Asymptotic behaviors for the calculated dispersions near the quadruple WPs, which is cubic in the (111) direction and quadratic in other directions.

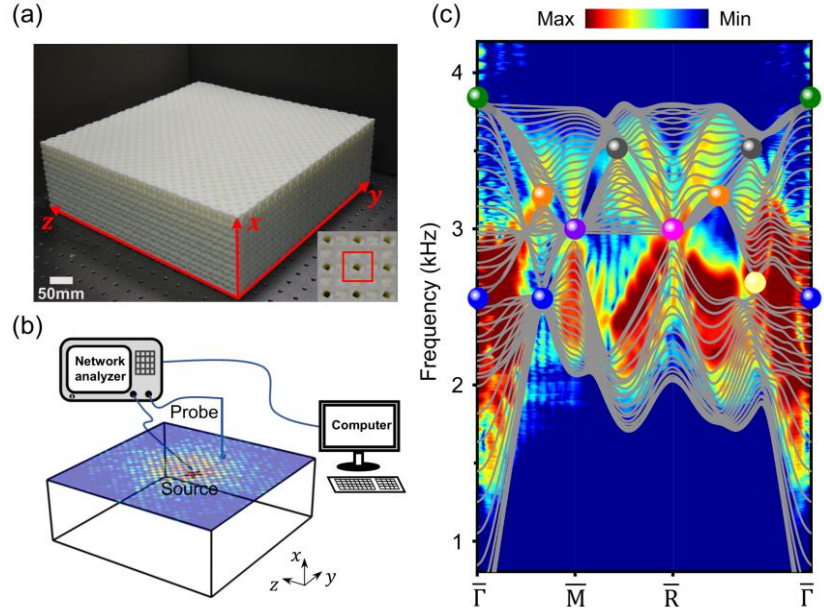

FIG. 2. (a) Photograph of the PC sample and the enlarged view of a unit cell on the surface boundary (boxed by red lines). has $6 \times 21 \times 21$ unit cells along the $x$, $y$, and $z$ directions. (b) Experimental setup. The color map denotes the acoustic field distribution at $3.50 \text{ kHz}$ and the red star represents the acoustic source. (c) Projected bulk dispersions along the high symmetry lines of the projected BZ. The color map and curves represent the experimental and numerical results, respectively. The colored balls denote the projections of the WPs.

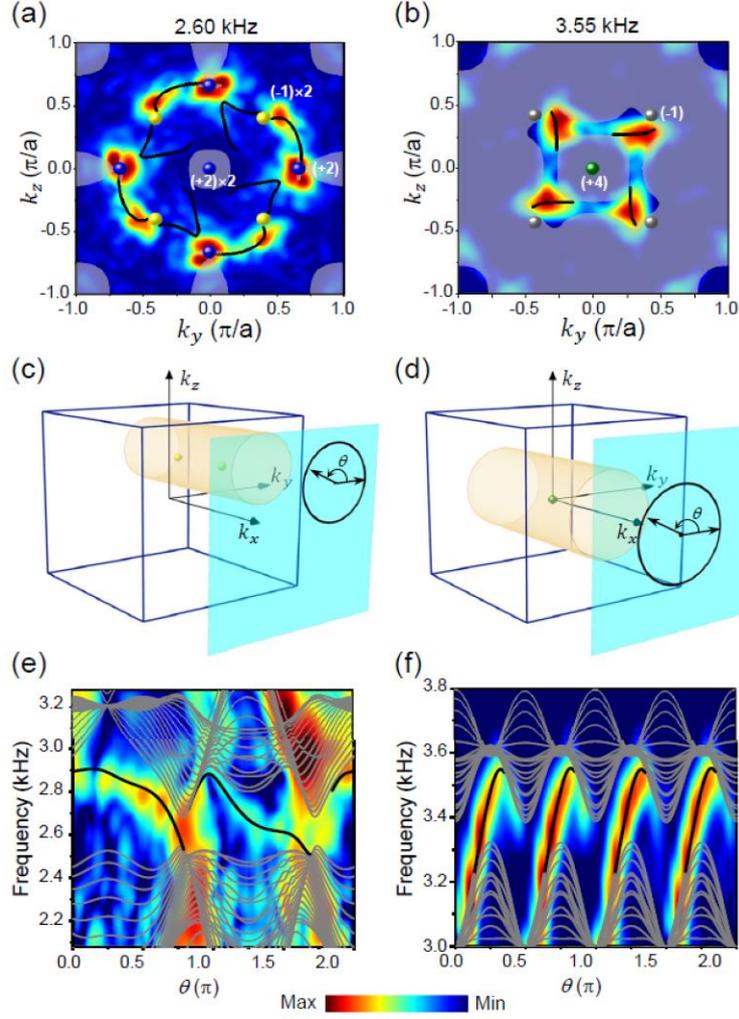

FIG. 3. (a) and (b) Surface state dispersions at the frequencies of 2.60 kHz and 3.55 kHz, respectively. The light gray areas denote the projected dispersions of the bulk states. The colored balls represent the projections of the WPs on the $k_y$-$k_z$ plane. (c) and (d) Cylinders in the 3D BZ with the baselines on the projected BZs. The radii of the black circles in (c) and (d) are $0.3\pi/a$ and $0.5\pi/a$, respectively. (e) and (f) Projected dispersions (right panels) along the circles with respect to $\theta$. Color maps represent experimental results and curves correspond to simulated results.

| Mark | Frequency | $E/t_0$ | Location in reduced BZ in unit of $\pi/a$ | Multiplicity | Bands | Band charges |
|---|---|---|---|---|---|---|
| 🔵 | 2.55 kHz | $-2$ | $(1/3, 0, 0)$ | 6 | 1 and 2 | 2 and $-2$ |
| 🟡 | 2.65 kHz | $-1.55$ | $(0.404, 0.404, 0.404)$ | 8 | 1 and 2 | $-1$ and 1 |
| 🟣 | 2.99 kHz | 0 | $(1, 1, 1)$ | 1 | 1, 2, and 3 | 2, 0, and $-2$ |
| 🟪 | 2.99 kHz | 0 | $(1, 1, 0)$ | 3 | 1, 2, and 3 | $-2$, 0, and 2 |
| 🟠 | 3.21 kHz | 1 | $(1/3, 1/3, 0)$ | 12 | 2 and 3 | 1 and $-1$ |
| ⚪ | 3.51 kHz | 2.44 | $(1, 0.430, 0.430)$ | 12 | 2 and 3 | $-1$ and 1 |
| 🟢 | 3.84 kHz | 4 | $(0, 0, 0)$ | 1 | 2 and 3 | 4 and $-4$ |

TABLE I. A summary on the WPs in the chiral PC, including the conventional WPs, the quadratic double WPs, the spin-1 WPs, and the quadruple WPs.